Fig. 1

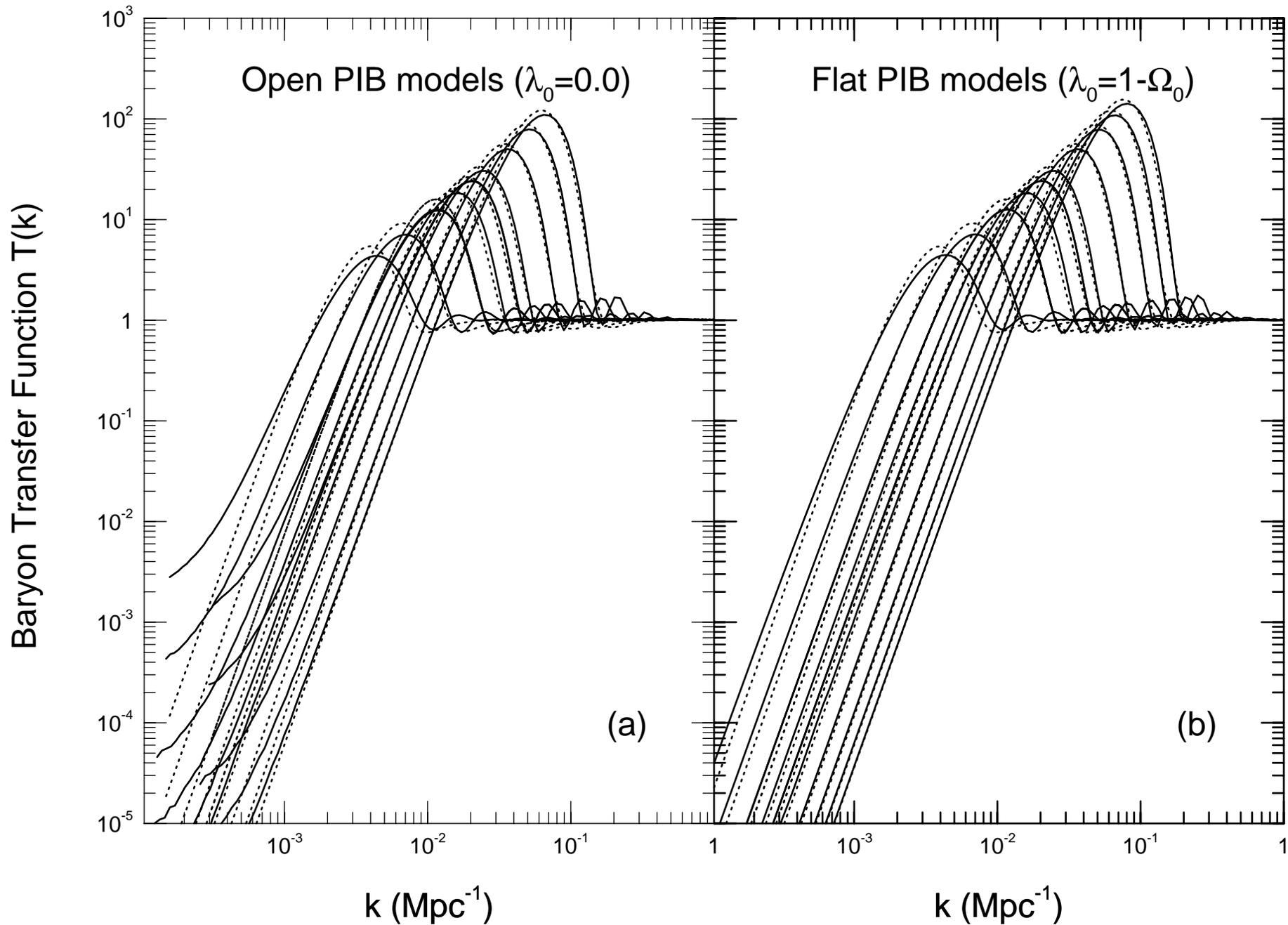

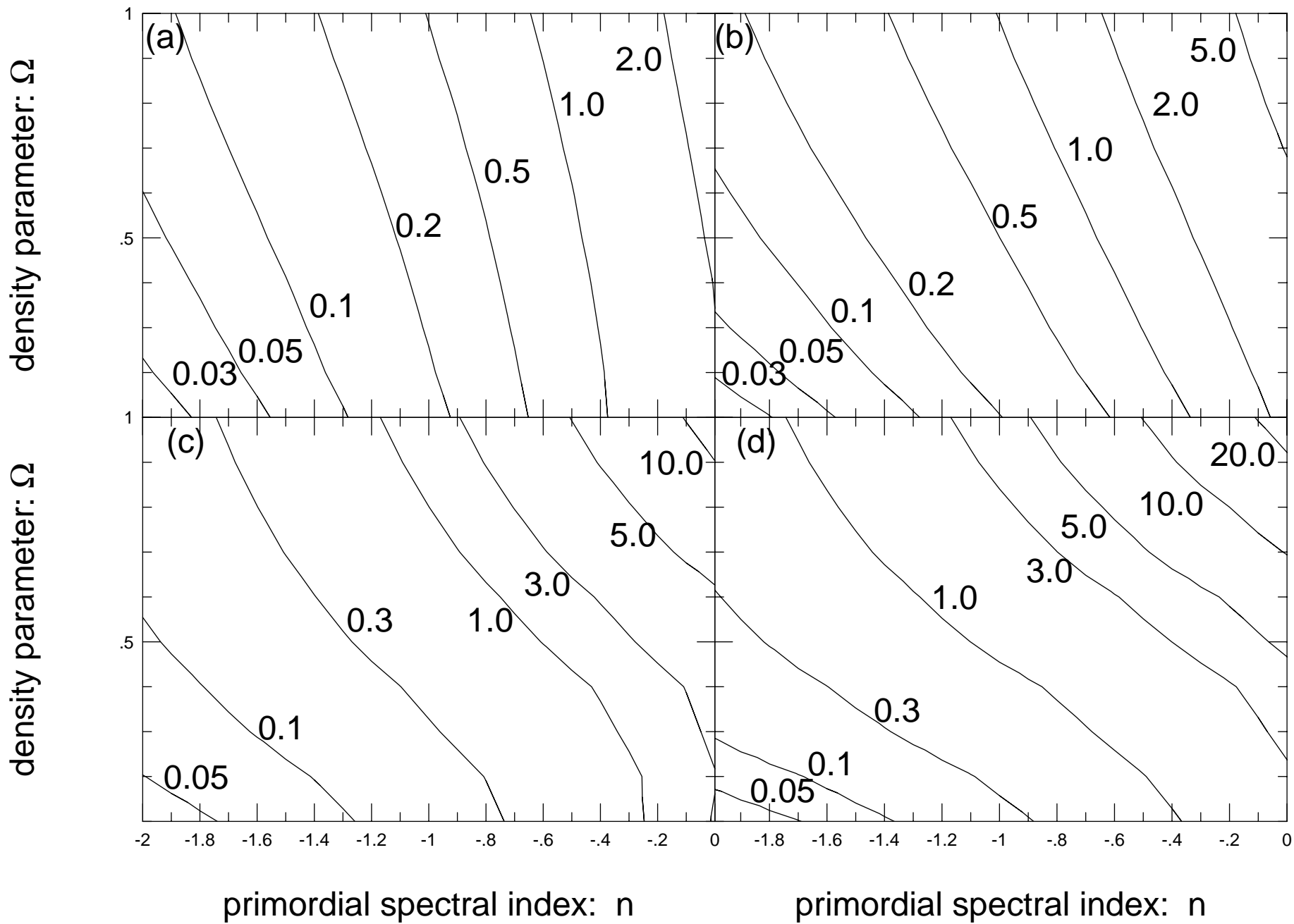

Fig. 2

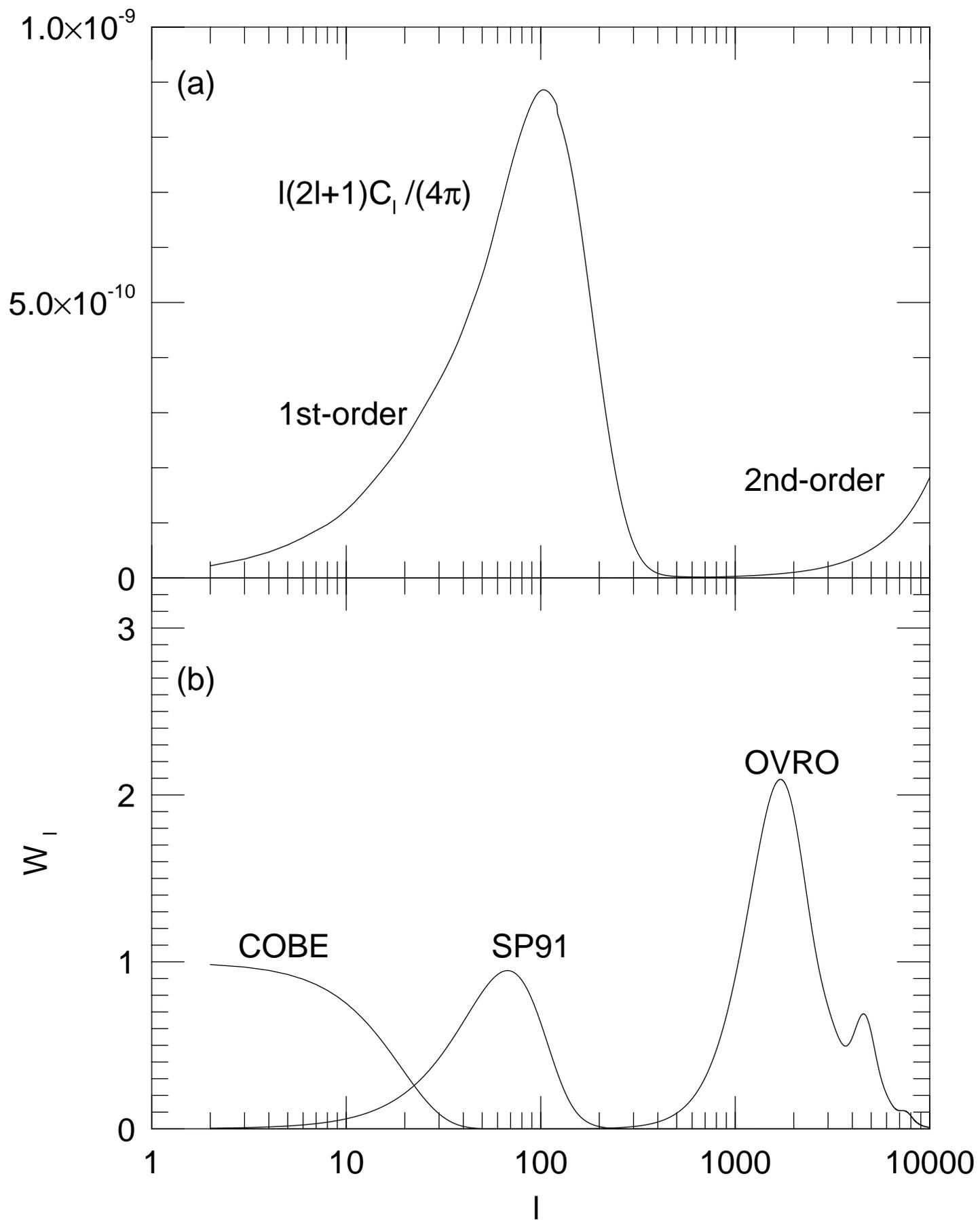

Fig. 3

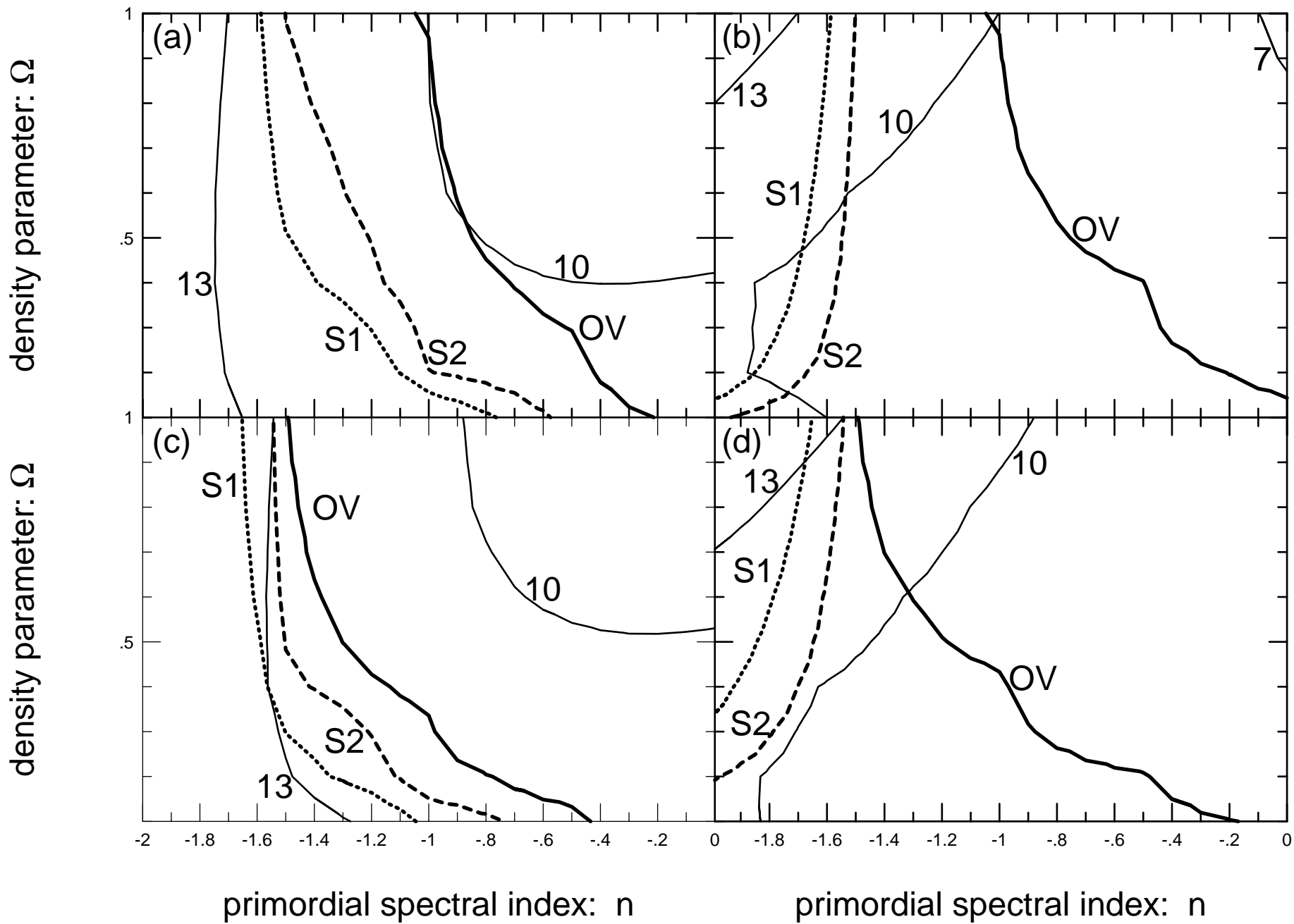

Fig. 4

# Microwave background anisotropies in primeval isocurvature baryon models: constraints on the cosmological parameters

Takashi Chiba [1], Naoshi Sugiyama [1,2] and Yasushi Suto [1,3] [†]

[1] *Department of Physics, The University of Tokyo, Bunkyo-ku, Tokyo 113, Japan*
[2] *Astronomy Department, University of California at Berkeley, Berkeley, CA 94720, USA*
[3] *Uji Research Center, Yukawa Institute for Theoretical Physics, Kyoto University Uji 611, Japan*

## Abstract

We have performed the most comprehensive predictions of the temperature fluctuations $\delta T/T$ in the primeval isocurvature baryon models to see whether or not the models are consistent with the recent data on the cosmic microwave background anisotropies. More specifically, we computed the $\delta T/T$ corresponding to the experimental set-up by the South-Pole and the Owens Valley experiments as well as the COBE satellite. The amplitudes of the predicted $\delta T/T$ are normalized by means of the COBE 10° data. The resulting constraints on the models are presented on $n - \Omega_b$ plane in the case of $\lambda_0 = 1 - \Omega_b$ (flat models) and $\lambda_0 = 0$ (open models), where $n$ is the primordial spectral index of entropy fluctuations and $\Omega_b$ is the present baryon density parameter. Our results imply that the PIB models cannot be reconciled with the current observations for any reasonable set of cosmological parameters.

*Subject headings*: cosmic microwave background—cosmology:theory

[†] e-mail addresses:tchiba@yayoi.phys.s.u-tokyo.ac.jp
sugiyama@bkyast.berkeley.edu
suto@yayoi.phys.s.u-tokyo.ac.jp

## 1. Introduction

The discovery of the temperature fluctuations $\delta T/T$ in the cosmic microwave background (CMB) by COBE (Smoot et al. 1992) strongly supports the gravitational instability picture for origin of hierarchical structures in the universe. In particular, the amplitude of $\delta T/T$ places tight constraints on the model parameters of viable scenarios (Wright et al. 1992; Gouda & Sugiyama 1992; Sugiyama & Gouda 1992). In the case of the standard cold dark matter (CDM) model in which the total density parameter $\Omega_0$ is unity, the dimensionless cosmological parameter $\lambda_0$ vanishes, and the dimensionless Hubble constant $h$ ($\equiv H_0/100 \text{km/sec/Mpc}$) is 0.5, for instance, the COBE $\delta T/T$ on $10°$ scale implies that the amplitude of top-hat mass fluctuations averaged over $r = 8h^{-1}\text{Mpc}$ sphere should be close to unity. It has been pointed out, however, that such unbiased standard CDM models have serious troubles in explaining the galaxy-galaxy two-point correlation function, and small-scale pair-wise velocity dispersions of galaxies among others (Davis et al. 1985; Suto & Suginohara 1991; Ueda, Itoh & Suto 1993; Suto 1993). In the present paper, we will re-examine the viability of yet another model of structure formation, primordial isocurvature baryon (PIB) models (Peebles 1987a,b), in the light of the COBE (Smoot et al. 1992), the recent ACME-HEMT South-Pole (SP : Gaier et al. 1992; Schuster et al. 1993) and Owens Valley (OVRO : Readhead et al. 1989) data.

There are a number of previous studies considering the astrophysical consequences of, and the resulting constraints on the PIB models(Peebles 1987a,b; Efstathiou & Bond 1987; Efstathiou 1988). They indicate that successful PIB models require a steep spectrum of primordial entropy (or baryon number) fluctuations; if the spectrum of the primordial entropy $S(k)$ ($k$ is the comoving wavenumber) is given by a simple power-law $\langle |S(k)|^2 \rangle \propto k^n$, then $n \sim -1$ is preferred rather than $n = -3$ predicted by conventional inflation models (Gouda, Sasaki & Suto 1989; Suginohara & Suto 1992; Cen, Ostriker & Peebles 1993). It turns out, however, that this contrived spectrum can be realized in a model of baryogenesis in the context of inflation (Yokoyama & Suto 1991). Unless a significant fraction of the present universe is dominated by non-baryonic dark matter which does not interact with radiation field, small scale excessive $\delta T/T$ is inevitably produced both in adiabatic and isocurvature models (Gouda, Sasaki & Suto 1987, 1989). This is not problematic in PIB models in which reionization of the universe at high redshifts is expected to occur due to the considerable power of fluctuations on small-scales. This early reionization, however, would produce a secondary $\delta T/T$ via the Doppler effect of the ionized plasma (Ostriker & Vishniac 1986; Vishniac 1987).

Detailed examination of the CMB anisotropies incorporating the above velocity-induced temperature fluctuations in the PIB models was made by Efstathiou and Bond (1987), and Efstathiou (1988). In the present work, we solve the Boltzmann equation directly up to the present epoch taking full account of all the first-order effects without approximation. Then we separately compute the contribution from the second-order velocity terms discussed by Vishniac (1987) which is added in quadrature to the direct numerical results to the first order. In doing so, we properly evaluate the growth rate of the second-order terms in $\Omega_0 < 1$ universes. With the above scheme, we compute the $\delta T/T$ which should be compared with the COBE $10°$ scale and quadrupole data, the SP experiments, and the OVRO experiment. The resulting constraints on the models are presented on $n - \Omega_b$ plane (in the present baryon-dominated models, $\Omega_0 = \Omega_b$ ) in the case of $\lambda_0 = 1 - \Omega_b$ (flat models) and $\lambda_0 = 0$ (open models). The former has not been studied quantitatively in the PIB scenarios so far.

## 2. Scenarios and parameters in PIB models

### 2.1. Basic scenario

In the present model, the primordial baryon number fluctuations are responsible for the formation of cosmic structures at much later epochs. The spectrum shape at the initial epoch $a_i$ when the fluctuations were generated is largely model-dependent and we simply adopt a single power-law form for the primordial entropy power spectrum $\langle |S(k)|^2 \rangle \propto k^n$ which is a reasonable approximation to the proposed models (Yokoyama & Suto 1991; Sasaki & Yokoyama 1991) on scales of astrophysical interest. As mentioned earlier large-scale galaxy distribution is consistent with $n = -1$ (Suginohara & Suto 1992; Cen et al. 1993) while conventional inflationary models predict $n = -3$. Since the universe is assumed to be dominated by baryons only, i.e., $\Omega_0 = \Omega_b$, Silk damping (Silk 1968) would smear out the small-scale density fluctuations if they were adiabatic. This is why one has to resort the isocurvature fluctuations for a realistic reionization scenario, even if the mode seems somewhat contrived. The details of the reionization history are quite uncertain or at least model-dependent, and thus we assume that the universe never recombined for simplicity. This corresponds to the case of the maximum erasure of the primary $\delta T/T$.

If the ionization rate of the hydrogen $x_e$ has been constant since the redshift $z$, the corresponding optical depth for the electron scattering is expressed as:

$$\tau(z) = \begin{cases} \dfrac{2\tau_0}{3\Omega_0^2} \left[ 2 - 3\Omega_0 + (\Omega_0 z + 3\Omega_0 - 2)\sqrt{1+\Omega_0 z} \right] & \text{(open models)} \\ \dfrac{2\tau_0}{3\Omega_0} \left[ \sqrt{\Omega_0(1+z)^3 + 1 - \Omega_0} - 1 \right] & \text{(flat models)} \end{cases} \quad (2.1)$$

where $\tau_0$ is the optical depth in the present horizon size ($= c/H_0$) and is given by

$$\tau_0 \equiv \frac{3c\,\sigma_T H_0}{8\pi G m_p} x_e \Omega_b \left(1 - \frac{Y_{He}}{2}\right) \sim 0.06 x_e \Omega_b h. \quad (2.2)$$

In the above expression, $c$, $\sigma_T$, $G$, and $m_p$ denote the velocity of light, the Thomson cross section, the gravitational constant, and the proton mass, respectively. In numerical evaluation of $\tau_0$, we adopt $Y_{He} = 0.23$ as the abundance of helium by mass. The epoch when $\tau(z)$ reaches unity is most important in discussing the effect of reionization. If the redshift of the epoch $z_c$ is much larger than $1/\Omega_0$, one may derive the following approximate expression for both open and flat models:

$$z_c \sim 30 \left(\frac{\Omega_0}{0.1}\right)^{1/3} \left(\frac{0.05}{x_e \Omega_b h}\right)^{2/3}. \quad (2.3)$$

In what follows, all quantities defined at this epoch $z_c$ are indicated by the subscript $c$. For example, $t_c$ denotes the cosmic time corresponding to the redshift $z_c$. Clearly almost complete ionization ($x_e = 1$) is necessary for the past universe to be opaque with respect to Thomson scattering.

Early reionization of the universe $z \gtrsim z_c$ would substantially smooth out the intrinsic anisotropies in $\delta T/T$, but in turn produce a significant amount of secondary anisotropies due to the Doppler and the inverse Compton effects (Zel'dovich & Sunyaev 1969) since some initial time $t_i$ up to the present $t_0$:

$$\frac{\delta T}{T} = \int_{t_i}^{t_0} \boldsymbol{\gamma} \cdot \mathbf{v} n_e \sigma_T dt, \quad (2.4)$$

$$\frac{\delta T}{T} = -2 \int_{t_i}^{t_0} \frac{k_B T_e}{m_e c^2} n_e \sigma_T c \, dt, \tag{2.5}$$

where $k_B$ is the Boltzmann constant, $T_e$, $m_e$, and $n_e$ are the temperature, mass and number density of electrons, and $\gamma$ is the direction of line-of-sight.

The $\delta T/T$ due to the cumulative Sunyaev-Zel'dovich effect (eq.[2.5]) from nonlinear objects at high redshifts was considered by Makino and Suto (1993). Although their results indicate that the effect potentially induces the observable amount of $\delta T/T$, the quantitative values are dependent on the assumption of the number density and the spatial correlation of the nonlinear objects at high redshifts. Therefore we focus our attention on the velocity-induced term (2.4) in what follows. Ostriker and Vishniac (1986) for the first time pointed out the possibility that the second-order velocity induced terms in eq.(2.4) might dominate the temperature fluctuations in some cosmological scenarios on small scale. Vishniac (1987) worked out that idea in detail and computed the $\delta T/T$ in CDM models with early reionization. Subsequently, this method was refined and applied to the PIB models ($\lambda_0 = 0$) by Efstathiou and Bond (1987). Recently, Hu, Scott and Silk (1993) and Dodelson and Jubas (1993) evaluated the magnitudes of terms contributed to the $\delta T/T$ up to the second-order, and concluded that the second-order velocity term considered by Vishniac (1987) gives the dominant contribution on small scales.

### 2.2. *Transfer function of baryons*

The evolution of isocurvature fluctuations is completely different from their adiabatic counterpart. For quantitative comparison of the $\delta T/T$ predicted in the PIB models and the observational data, one has to numerically solve the linearized Boltzmann equations which describe the general relativistic evolution of photons, baryons, and massless neutrinos. Nevertheless it is instructive to summarize beautiful analytical results based on perturbation theory by Kodama and Sasaki (1986), who considered baryon-photon universes neglecting the contribution of neutrinos. They showed that the entropy fluctuations $S(k)$ is nearly constant in isocurvature models, and that the spectrum of total energy density fluctuations after the matter-radiation equality epoch reduce to

$$\langle |\delta_t(k)|^2 \rangle \sim \begin{cases} \langle |S(k)|^2 \rangle & (k \gg a_{eq} H_{eq}) \\ \dfrac{4}{225} \left(\dfrac{a}{a_0}\right)^2 \dfrac{k^4}{\Omega_b^2 (a_0 H_0)^4} \langle |S(k)|^2 \rangle & (k \ll a_{eq} H_{eq}) \end{cases} \tag{2.6}$$

where $a$ is the scale factor at that epoch, and $a_{eq}$ and $a_0$ are those at the equality and the present epochs, respectively. Therefore even if the primordial entropy fluctuations $\langle |S(k)|^2 \rangle$ are given by a scale-free form $\propto k^n$, the resulting total fluctuations at $a > a_{eq}$ develop a peak near a comoving scale for the Hubble length at $a_{eq}$:

$$\lambda_{peak} \equiv \frac{2\pi}{a_{eq} H_{eq} (15/8\sqrt{2})} \sim 45 (\Omega_b h^2)^{-1} \mathrm{Mpc}. \tag{2.7}$$

Since the present spectrum of baryons would correspond to eq.(2.6) at the decoupling epoch of baryons and photons, $z_c$, the height of the peak is characterized by the ratio of the power spectrum at each side:

$$R \equiv \frac{4}{225 \Omega_b^2} z_c^{-2} \left(\frac{15}{8\sqrt{2}} \frac{a_{eq} H_{eq}}{a_0 H_0}\right)^4 \sim 40 \left(\frac{\Omega_b h^2}{0.01}\right)^2 \left(\frac{30}{z_c}\right)^2. \tag{2.8}$$

Therefore as $\Omega_b h^2$ increases, the more prominent peak forms at the smaller scale. The peak height becomes larger if the decoupling occurs late.

The above features should be compared with the direct numerical computation. In Figure 1 are plotted the transfer functions of the baryon density fluctuations $T(k)$ up to the present epoch $a_0$:

$$T(k) \equiv \frac{\langle|\delta_b(k, a_0)|^2\rangle}{\langle|S(k)|^2\rangle}. \tag{2.9}$$

Figure 1a displays the results for open models, while Figure 1b for flat models. Curves in each panel correspond to $\Omega_b = 0.1, 0.2, 0.4, 0.6, 0.8$ and $1.0$ from top to bottom, both for $h = 0.5$ and $1.0$. Since this spectrum plays a fundamental role in computing the velocity induced $\delta T/T$ below, we fitted the numerical results (solid curves) by the following form (dotted curves):

$$T(k) = [\hat{a}k^2 \exp\left(-\hat{b}k^2\right) + \exp\left(-\frac{\hat{c}}{k^2} + \frac{\hat{d}}{k}\right)]^2, \tag{2.10}$$

with

$$\hat{a} = \frac{6000}{(\Omega_b h^2)^{1.16}} \mathrm{Mpc}^2, \tag{2.11}$$

$$\hat{b} = \begin{cases} \dfrac{180}{(\Omega_b h^2)^{1.72}} \mathrm{Mpc}^2 & (\Omega_b h^2 > 0.15), \\ \dfrac{220}{(\Omega_b h^2)^{1.58}} \mathrm{Mpc}^2 & (\Omega_b h^2 < 0.15), \end{cases} \tag{2.12}$$

$$\hat{c} = 0.01 \left(\Omega_b h^2\right)^{1.6} \mathrm{Mpc}^{-2}, \tag{2.13}$$

$$\hat{d} = 0.02 \left(\Omega_b h^2\right)^{0.8} \mathrm{Mpc}^{-1}. \tag{2.14}$$

The difference between $\lambda_0 = 0$ and $\lambda = 1 - \Omega_b$ models manifests only for very small $k$ (Fig.1) which gives only negligible contribution to the resulting $\delta T/T$ below. Thus we use the same fits in both cases.

In our present models, there are four free parameters, i.e., the primordial power-law index of entropy ( or baryon number) fluctuations $n$, the baryon density parameter $\Omega_b$ ($= \Omega_0$ in the present models), the dimensionless cosmological constant $\lambda_0$, and finally the dimensionless Hubble constant $h$. More precisely, we adopt $\langle|S(k)|^2\rangle \propto \tilde{k}^n$, rather than $\propto k^n$, where $\tilde{k} \equiv \sqrt{k^2 + K}$ with $K$ being the curvature constant. In particular, we explore the constraints on $n - \Omega_b$ plane for $0.1 \leq \Omega_b \leq 1$ and $-2 \leq n \leq 0$. As for the remaining two parameters, we consider typical four sets, i.e.,$(\lambda_0, h) = (0, 0.5), (0, 1.0), (1 - \Omega_b, 0.5)$, and $(1 - \Omega_b, 1.0)$.

## 3. Temperature anisotropies

### 3.1. Method of computing temperature fluctuations

The linear part of $\delta T/T$ in the microwave background is obtained by numerical integration of the linearized Boltzmann equations for photons, baryons, and massless neutrinos on the basis of the gauge-invariant formalism (see, e.g., Sugiyama & Gouda 1992). The temperature anisotropy $\Delta_T$ is expanded in Fourier space by writing:

$$\Delta_T(\eta, \mu, k) = \sum_\ell (-i)^\ell (2\ell + 1) a_\ell(\eta, k) P_\ell(\mu), \tag{3.1}$$

where $\eta$ is conformal time, $\mu$ is direction cosine and $P_\ell$ is the Legendre polynomial of the $\ell$-th order. Then the Boltzmann equation is reduced to recursive equations of $a_\ell$. We do not use a free-streaming approximation but rather directly solve the equations until the present epoch $\eta_0$. Hence our computation fully includes all linear terms such as the Doppler (first-order velocity) effect, the Sachs-Wolfe effect and the line of sight integration of the derivative of the gravitational potential. Therefore our remaining task is simply to add the second-order velocity-induced terms to the primary $\delta T/T$ so computed. In doing so, we basically follow the prescription described by Efstathiou (1988) except that we treat the growth rate in $\Omega_0 < 1$ universes more properly. We summarize the final expressions below following his notation.

The velocity induced temperature fluctuations can be evaluated by solving the radiation transfer equations perturbatively. The second-order velocity induced part of the direction averaged power-spectrum of the temperature fluctuations $W_T^{v2}$ is expressed as

$$\begin{aligned}
W_T^{v2} &\equiv \frac{1}{2}\int_{-1}^{1} |\Delta_T^{v2}(\eta_0, \mu, k)|^2 d\mu \tag{3.2} \\
&= \frac{4V_x P^2(k, \eta_c)}{\pi \eta_c^3} I_1 I_2(k), \\
I_1 &= \int_0^{\eta_0} \frac{1}{4}\left[\frac{\dot{D}(\eta')}{D(\eta_c)}\frac{D(\eta')}{D(\eta_c)}\right]^2 [g(\eta_0, \eta')\eta_c]^2 \eta_c^2 d(\frac{\eta}{\eta_c}), \tag{3.3} \\
I_2(k) &= \int_0^\infty \int_{-1}^1 dy d\mu \frac{(1-\mu^2)(1-2\mu y)^2}{(1+y^2-2\mu y)^2}\frac{P\left[k\sqrt{1+y^2-2\mu y}, \eta_c\right]}{P(k, \eta_c)}\frac{P(ky, \eta_c)}{P(k, \eta_c)}. \tag{3.4}
\end{aligned}$$

where the suffix $c$ represents the time at $z_c$, $D(\eta)$ is the linear growth rate of matter density fluctuations, and

$$g(\eta, \eta') \equiv [\sigma_T \bar{n}_e a]_{\eta'} \exp\left[-\int_{\eta'}^\eta \sigma_T \bar{n}_e a \, d\eta\right]. \tag{3.5}$$

The previous work by Efstathiou and Bond (1987) and Efstathiou (1988) approximated the numerical factors $I_1$ as 6.38 irrespectively of the underlying cosmology. In the following computation, we numerically evaluate $I_1$ taking account of the dependence on $\Omega_0$ and $\lambda_0$. For instance, $I_1$ becomes 1.68 (3.97) for $h = 0.5, \Omega_0 = 0.1$ and $\lambda_0 = 0$ (0.9) . Thus the overall amplitude of the velocity induced $\delta T/T$ differs by a factor of 1.95 (1.27).

As a check of our code, we have compared our numerical results with the available data computed by other investigators and found good agreement; with Efstathiou and Bond (1987), Peebles (1993) and Stompor (1993) for transfer function, and with Efstathiou and Bond (1987) for CMB power spectra.

### 3.2. Constraints on $\Omega_b$ and $n$

In order to make a direct comparison between the CMB power spectrum and specific observations, we introduce the coefficients of the CMB anisotropies in the $\ell$-mode defined as

$$C_\ell \equiv \frac{2}{\pi} \int_0^\infty d\tilde{k}\tilde{k}^2 \frac{\left(\tilde{k}^2 - K\right) \cdots \left(\tilde{k}^2 - K\,\ell^2\right)}{\left(\tilde{k}^2 - K\right)^\ell} |a_\ell|^2. \quad (3.6)$$

The expected temperature anisotropy for each experiment is expressed in terms of $C_\ell$ and the specific window function $W_\ell$:

$$\left(\frac{\Delta T}{T}\right)^2_{exp} = \sum_{\ell \geq 2} \frac{2\ell+1}{4\pi} C_\ell W_\ell. \quad (3.7)$$

We normalize the fluctuation amplitude by means of the COBE $\delta T/T$ at $10°$ scale (Smoot et al. 1992):

$$\Delta T = (30 \pm 5)\mu K, \qquad \frac{\Delta T}{T} = (1.1 \pm 0.2) \times 10^{-5}. \quad (3.8)$$

First of all, in order to exhibit the resulting amplitude of the baryon density fluctuations in these models, we display the contour plots of the root-mean-square mass fluctuation $\sigma_8$ at $r = 8h^{-1}$Mpc for a top-hat sphere (Fig.2). As is clearly indicated in Figure 2, $\sigma_8$ is very sensitive to the spectral index $n$.

Next, with the above amplitude normalization, we predict the temperature anisotropies expected in specific CMB experiments. We consider the COBE quadrupole (Smoot et al. 1992), OVRO (Readhead et al. 1989), 9 points South Pole (SP91 : Gaier et al. 1992) and 13 points South Pole (SP93 : Schuster et al. 1993) experiments. As for the SP91 experiments, we only consider the highest frequency channel because the data in the other channels might be contaminated by foreground galactic emission (Gaier et al. 1992). The window function $W_\ell$ of each experiment and $C_\ell$ of the model with $\Omega_b = 0.2$, $h = 0.5$ and $n = 0$ are shown in Figure 3. The contribution of the second-order velocity term is also presented. Apparently, we expect large temperature fluctuations on the scale of SP experiments for this model. The second-order velocity term is dominated and only important on OVRO scale. In Figure 4 are shown the contour plots for the COBE quadrupole, OVRO, SP91 and SP93 experiments on $n - \Omega_b$ plane. The value of the quadrupole anisotropy detected by COBE is $(13 \pm 4)\mu K$. Our expected quadrupole anisotropies for desirable models, i.e., $\Omega_b = (0.1 \sim 0.3)$, and $n = (-1 \sim 0)$ are typically smaller than this value. If we take into account *cosmic variance*, however, these expected values are still consistent with observed one. To obtain constraints from OVRO and SP experiments, we use Bayesian method (Bond et al. 1991) and assume a uniform prior probability density. The limits from the OVRO experiment is generally weaker than the constraints from the SP data. It is interesting to note that the completely independent experiments with different instruments and the effective scales of observation consistently point to $n \lesssim -1.5$ as an allowed region. The PIB models with $\lambda_0 = 0$ require that $\Omega_b \gtrsim 0.5$ additionally, which is in serious conflict with the prediction of the standard big-bang nucleosynthesis (SBBN), i.e., $\Omega_b h^2 = (0.01 \sim 0.02)$. With nonvanishing $\lambda_0$, especially $\lambda_0 = 1 - \Omega_b$ as we consider here, the PIB models are consistent with the SBBN prediction if $n \lesssim -1$. The problem with such models, however, is that they predict very low values of $\sigma_8$ implying that visible galaxies should be significantly biased with respect to the underlying mass distribution. Since the PIB models are motivated by the assumption that non-baryonic dark matter does not exist, such conclusions seem quite contrived even if not excluded.

## 4. Conclusions and discussion

We have computed the $\delta T/T$ in the PIB models to see whether or not the recent data including COBE, SP, and OVRO observations can be reconciled in a consistent set of cosmological parameters. As we have shown in the above, the present results constrain the range of parameters in the PIB models very stringently, and the models should be excluded. This conclusion, however, is subject to change by considering the following possibilities; the first is the uncertainty due to the cosmic variance. This is especially important in interpreting the quadrupole amplitude from the COBE data. In addition, a relatively small sky coverage of the ground-based experiments including SP91 and SP93 might result in statistical ambiguity to some extent (Bunn et al. 1993). A clear resolution to the problem requires future observations with independent and large sky coverage and possibly at multi-wavebands. Second, we assumed in the present analysis for simplicity that the universe never recombined. Apparently this is not realistic. This assumption results in the most significant erasure of the primary $\delta T/T$, but at the same time the magnitude of the velocity-induced $\delta T/T$ becomes largest. Therefore it is not clear whether the overall prediction for the $\delta T/T$ increases or decreases when more realistic thermal history is considered. This should be worked out in details with a model of reionization after the standard recombination epoch.

The authors would like to thank N. Gouda, W. Hu, D. Scott, and J. Silk for valuable discussions. N.S. gratefully acknowledges financial support from a JSPS (the Japan Society of the Promotion of Sciences) postdoctoral fellowship for research abroad. The present computations were carried out on VP2600 at the Data Processing Center at Kyoto University, and on M780 at National Astronomical Observatory, Japan. This research was supported in part by the Grants-in-Aid by the Ministry of Education, Science and Culture of Japan (03302012, 04352006, 04740130, and 05640312).

T. Chiba, N. Sugiyama, & Y. Suto                                                                                9

**FIGURE CAPTIONS**

**Figure 1** : Transfer functions of baryon density fluctuations. (a) open models ($\lambda_0 = 0$); (b) flat models ($\lambda_0 = 1 - \Omega_b$).

**Figure 2** : Contour map of $\sigma_8$ normalized by the COBE 10° result. (a) h=0.5, $\lambda_0 = 1 - \Omega_b$; (b) h=0.5, $\lambda_0 = 0$; (c) h=1.0, $\lambda_0 = 1 - \Omega_b$; (d) h=1.0, $\lambda_0 = 0$.

**Figure 3** : (a)Power spectrum of temperature anisotropies $\ell(2\ell+1)C_\ell/4\pi$ for the model in which $\Omega_b = 0.2$, $h = 0.5$ and $n = 0$. (b)Window functions $W_\ell$ for the COBE, SP91 and OVRO experiments.

**Figure 4** : Constrains on $n$ and $\Omega_b$ from OVRO, SP91, SP93 and COBE quadrupole anisotropy normalized by the COBE 10° result. (a) h=0.5, $\lambda_0 = 1 - \Omega_b$; (b) h=0.5, $\lambda_0 = 0$; (c) h=1.0, $\lambda_0 = 1 - \Omega_b$; (d) h=1.0, $\lambda_0 = 0$. OV(thick solid lines), S1(thick dotted lines) and S2(thick short-dashed lines) are the 95% confidence-limit from OVRO, SP91 and SP93 experiments, respectively. Regions to the right of each curve are excluded by the corresponding experimental results. Solid lines labeled as 7, 10 and 13 are the countour curves for the COBE quadrupole anisotropies in units of $\mu$K.